\documentclass[conference]{IEEEtran}
\IEEEoverridecommandlockouts
\usepackage{cite}
\usepackage{amsmath,amssymb,amsfonts}
\usepackage{algorithmic}
\usepackage{graphicx}
\usepackage{textcomp}
\usepackage{xcolor}
\usepackage{amsmath,graphicx}
\usepackage{tcolorbox}
\usepackage[draft]{minted}
\definecolor{codeBg}{HTML}{FFFFFF}  
\usepackage{booktabs,siunitx}

\newcommand{\splitcell}[1]{%
  \begingroup
  \renewcommand{\arraystretch}{1}%
  \begin{tabular}{@{}c@{}}#1\end{tabular}%
  \endgroup
}

\setminted[python]{
    breaklines=true,
    encoding=utf8,
    fontsize=\footnotesize,
    bgcolor=codeBg
}
\makeatletter
\def\PYG@reset{\let\PYG@it=\relax \let\PYG@bf=\relax%
    \let\PYG@ul=\relax \let\PYG@tc=\relax%
    \let\PYG@bc=\relax \let\PYG@ff=\relax}
\def\PYG@tok#1{\csname PYG@tok@#1\endcsname}
\def\PYG@toks#1+{\ifx\relax#1\empty\else%
    \PYG@tok{#1}\expandafter\PYG@toks\fi}
\def\PYG@do#1{\PYG@bc{\PYG@tc{\PYG@ul{%
    \PYG@it{\PYG@bf{\PYG@ff{#1}}}}}}}
\def\PYG#1#2{\PYG@reset\PYG@toks#1+\relax+\PYG@do{#2}}

\@namedef{PYG@tok@w}{\def\PYG@tc##1{\textcolor[rgb]{0.73,0.73,0.73}{##1}}}
\@namedef{PYG@tok@c}{\let\PYG@it=\textit\def\PYG@tc##1{\textcolor[rgb]{0.24,0.48,0.48}{##1}}}
\@namedef{PYG@tok@cp}{\def\PYG@tc##1{\textcolor[rgb]{0.61,0.40,0.00}{##1}}}
\@namedef{PYG@tok@k}{\let\PYG@bf=\textbf\def\PYG@tc##1{\textcolor[rgb]{0.00,0.50,0.00}{##1}}}
\@namedef{PYG@tok@kp}{\def\PYG@tc##1{\textcolor[rgb]{0.00,0.50,0.00}{##1}}}
\@namedef{PYG@tok@kt}{\def\PYG@tc##1{\textcolor[rgb]{0.69,0.00,0.25}{##1}}}
\@namedef{PYG@tok@o}{\def\PYG@tc##1{\textcolor[rgb]{0.40,0.40,0.40}{##1}}}
\@namedef{PYG@tok@ow}{\let\PYG@bf=\textbf\def\PYG@tc##1{\textcolor[rgb]{0.67,0.13,1.00}{##1}}}
\@namedef{PYG@tok@nb}{\def\PYG@tc##1{\textcolor[rgb]{0.00,0.50,0.00}{##1}}}
\@namedef{PYG@tok@nf}{\def\PYG@tc##1{\textcolor[rgb]{0.00,0.00,1.00}{##1}}}
\@namedef{PYG@tok@nc}{\let\PYG@bf=\textbf\def\PYG@tc##1{\textcolor[rgb]{0.00,0.00,1.00}{##1}}}
\@namedef{PYG@tok@nn}{\let\PYG@bf=\textbf\def\PYG@tc##1{\textcolor[rgb]{0.00,0.00,1.00}{##1}}}
\@namedef{PYG@tok@ne}{\let\PYG@bf=\textbf\def\PYG@tc##1{\textcolor[rgb]{0.80,0.25,0.22}{##1}}}
\@namedef{PYG@tok@nv}{\def\PYG@tc##1{\textcolor[rgb]{0.10,0.09,0.49}{##1}}}
\@namedef{PYG@tok@no}{\def\PYG@tc##1{\textcolor[rgb]{0.53,0.00,0.00}{##1}}}
\@namedef{PYG@tok@nl}{\def\PYG@tc##1{\textcolor[rgb]{0.46,0.46,0.00}{##1}}}
\@namedef{PYG@tok@ni}{\let\PYG@bf=\textbf\def\PYG@tc##1{\textcolor[rgb]{0.44,0.44,0.44}{##1}}}
\@namedef{PYG@tok@na}{\def\PYG@tc##1{\textcolor[rgb]{0.41,0.47,0.13}{##1}}}
\@namedef{PYG@tok@nt}{\let\PYG@bf=\textbf\def\PYG@tc##1{\textcolor[rgb]{0.00,0.50,0.00}{##1}}}
\@namedef{PYG@tok@nd}{\def\PYG@tc##1{\textcolor[rgb]{0.67,0.13,1.00}{##1}}}
\@namedef{PYG@tok@s}{\def\PYG@tc##1{\textcolor[rgb]{0.73,0.13,0.13}{##1}}}
\@namedef{PYG@tok@sd}{\let\PYG@it=\textit\def\PYG@tc##1{\textcolor[rgb]{0.73,0.13,0.13}{##1}}}
\@namedef{PYG@tok@si}{\let\PYG@bf=\textbf\def\PYG@tc##1{\textcolor[rgb]{0.64,0.35,0.47}{##1}}}
\@namedef{PYG@tok@se}{\let\PYG@bf=\textbf\def\PYG@tc##1{\textcolor[rgb]{0.67,0.36,0.12}{##1}}}
\@namedef{PYG@tok@sr}{\def\PYG@tc##1{\textcolor[rgb]{0.64,0.35,0.47}{##1}}}
\@namedef{PYG@tok@ss}{\def\PYG@tc##1{\textcolor[rgb]{0.10,0.09,0.49}{##1}}}
\@namedef{PYG@tok@sx}{\def\PYG@tc##1{\textcolor[rgb]{0.00,0.50,0.00}{##1}}}
\@namedef{PYG@tok@m}{\def\PYG@tc##1{\textcolor[rgb]{0.40,0.40,0.40}{##1}}}
\@namedef{PYG@tok@gh}{\let\PYG@bf=\textbf\def\PYG@tc##1{\textcolor[rgb]{0.00,0.00,0.50}{##1}}}
\@namedef{PYG@tok@gu}{\let\PYG@bf=\textbf\def\PYG@tc##1{\textcolor[rgb]{0.50,0.00,0.50}{##1}}}
\@namedef{PYG@tok@gd}{\def\PYG@tc##1{\textcolor[rgb]{0.63,0.00,0.00}{##1}}}
\@namedef{PYG@tok@gi}{\def\PYG@tc##1{\textcolor[rgb]{0.00,0.52,0.00}{##1}}}
\@namedef{PYG@tok@gr}{\def\PYG@tc##1{\textcolor[rgb]{0.89,0.00,0.00}{##1}}}
\@namedef{PYG@tok@ge}{\let\PYG@it=\textit}
\@namedef{PYG@tok@gs}{\let\PYG@bf=\textbf}
\@namedef{PYG@tok@gp}{\let\PYG@bf=\textbf\def\PYG@tc##1{\textcolor[rgb]{0.00,0.00,0.50}{##1}}}
\@namedef{PYG@tok@go}{\def\PYG@tc##1{\textcolor[rgb]{0.44,0.44,0.44}{##1}}}
\@namedef{PYG@tok@gt}{\def\PYG@tc##1{\textcolor[rgb]{0.00,0.27,0.87}{##1}}}
\@namedef{PYG@tok@err}{\def\PYG@bc##1{{\setlength{\fboxsep}{\string -\fboxrule}\fcolorbox[rgb]{1.00,0.00,0.00}{1,1,1}{\strut ##1}}}}
\@namedef{PYG@tok@kc}{\let\PYG@bf=\textbf\def\PYG@tc##1{\textcolor[rgb]{0.00,0.50,0.00}{##1}}}
\@namedef{PYG@tok@kd}{\let\PYG@bf=\textbf\def\PYG@tc##1{\textcolor[rgb]{0.00,0.50,0.00}{##1}}}
\@namedef{PYG@tok@kn}{\let\PYG@bf=\textbf\def\PYG@tc##1{\textcolor[rgb]{0.00,0.50,0.00}{##1}}}
\@namedef{PYG@tok@kr}{\let\PYG@bf=\textbf\def\PYG@tc##1{\textcolor[rgb]{0.00,0.50,0.00}{##1}}}
\@namedef{PYG@tok@bp}{\def\PYG@tc##1{\textcolor[rgb]{0.00,0.50,0.00}{##1}}}
\@namedef{PYG@tok@fm}{\def\PYG@tc##1{\textcolor[rgb]{0.00,0.00,1.00}{##1}}}
\@namedef{PYG@tok@vc}{\def\PYG@tc##1{\textcolor[rgb]{0.10,0.09,0.49}{##1}}}
\@namedef{PYG@tok@vg}{\def\PYG@tc##1{\textcolor[rgb]{0.10,0.09,0.49}{##1}}}
\@namedef{PYG@tok@vi}{\def\PYG@tc##1{\textcolor[rgb]{0.10,0.09,0.49}{##1}}}
\@namedef{PYG@tok@vm}{\def\PYG@tc##1{\textcolor[rgb]{0.10,0.09,0.49}{##1}}}
\@namedef{PYG@tok@sa}{\def\PYG@tc##1{\textcolor[rgb]{0.73,0.13,0.13}{##1}}}
\@namedef{PYG@tok@sb}{\def\PYG@tc##1{\textcolor[rgb]{0.73,0.13,0.13}{##1}}}
\@namedef{PYG@tok@sc}{\def\PYG@tc##1{\textcolor[rgb]{0.73,0.13,0.13}{##1}}}
\@namedef{PYG@tok@dl}{\def\PYG@tc##1{\textcolor[rgb]{0.73,0.13,0.13}{##1}}}
\@namedef{PYG@tok@s2}{\def\PYG@tc##1{\textcolor[rgb]{0.73,0.13,0.13}{##1}}}
\@namedef{PYG@tok@sh}{\def\PYG@tc##1{\textcolor[rgb]{0.73,0.13,0.13}{##1}}}
\@namedef{PYG@tok@s1}{\def\PYG@tc##1{\textcolor[rgb]{0.73,0.13,0.13}{##1}}}
\@namedef{PYG@tok@mb}{\def\PYG@tc##1{\textcolor[rgb]{0.40,0.40,0.40}{##1}}}
\@namedef{PYG@tok@mf}{\def\PYG@tc##1{\textcolor[rgb]{0.40,0.40,0.40}{##1}}}
\@namedef{PYG@tok@mh}{\def\PYG@tc##1{\textcolor[rgb]{0.40,0.40,0.40}{##1}}}
\@namedef{PYG@tok@mi}{\def\PYG@tc##1{\textcolor[rgb]{0.40,0.40,0.40}{##1}}}
\@namedef{PYG@tok@il}{\def\PYG@tc##1{\textcolor[rgb]{0.40,0.40,0.40}{##1}}}
\@namedef{PYG@tok@mo}{\def\PYG@tc##1{\textcolor[rgb]{0.40,0.40,0.40}{##1}}}
\@namedef{PYG@tok@ch}{\let\PYG@it=\textit\def\PYG@tc##1{\textcolor[rgb]{0.24,0.48,0.48}{##1}}}
\@namedef{PYG@tok@cm}{\let\PYG@it=\textit\def\PYG@tc##1{\textcolor[rgb]{0.24,0.48,0.48}{##1}}}
\@namedef{PYG@tok@cpf}{\let\PYG@it=\textit\def\PYG@tc##1{\textcolor[rgb]{0.24,0.48,0.48}{##1}}}
\@namedef{PYG@tok@c1}{\let\PYG@it=\textit\def\PYG@tc##1{\textcolor[rgb]{0.24,0.48,0.48}{##1}}}
\@namedef{PYG@tok@cs}{\let\PYG@it=\textit\def\PYG@tc##1{\textcolor[rgb]{0.24,0.48,0.48}{##1}}}

\def\PYGZus{\char`\_}

\def\PYGZsh{\char`\#}

\def\PYGZhy{\char`\-}
\def\PYGZsq{\char`\'}

\makeatother

\makeatletter 
\newcommand{\linebreakand}{%
  \end{@IEEEauthorhalign}
  \hfill\mbox{}\par
  \mbox{}\hfill\begin{@IEEEauthorhalign}
}
\makeatother 

\def\BibTeX{{\rm B\kern-.05em{\sc i\kern-.025em b}\kern-.08em
    T\kern-.1667em\lower.7ex\hbox{E}\kern-.125emX}}
\begin{document}

\title{Wav2Small: Distilling Wav2Vec2 to 72K Parameters for Low-Resource Speech Emotion Recognition
\thanks{Funding from the Horizon Europe programme under the SHIFT project (Grant Agreement No. 101060660) is greatly acknowledged.}
}
\author{\IEEEauthorblockN{Dionyssos Kounadis Bastian}
\IEEEauthorblockA{\textit{audEERING GmbH}, Gilching, Germany \\
\textit{dkounadis@audeering.com}
}
\and
\IEEEauthorblockN{Oliver Schr\"{u}fer}
\IEEEauthorblockA{\textit{audEERING GmbH}, Gilching, Germany \\
\textit{oschrufer@audeering.com}
}
\and 
\IEEEauthorblockN{Anna Derington}
\IEEEauthorblockA{\textit{audEERING GmbH}, Gilching, Germany \\
\textit{aderington@audeering.com}
}
\linebreakand
\IEEEauthorblockN{Hagen Wierstorf}
\IEEEauthorblockA{\textit{audEERING GmbH}, Gilching, Germany \\
\textit{hwierstorf@audeering.com}
}
\and
\IEEEauthorblockN{Florian Eyben}
\IEEEauthorblockA{\textit{audEERING GmbH}, Gilching, Germany \\
\textit{fe@audeering.com}
}
\and
\IEEEauthorblockN{Felix Burkhardt}
\IEEEauthorblockA{\textit{audEERING GmbH}, Gilching, Germany \\
\textit{fburkhardt@audeering.com}
}
\linebreakand
\IEEEauthorblockN{Björn W. Schuller}
\IEEEauthorblockA{\textit{audEERING GmbH}, Gilching, Germany, \\
\textit{CHI - Chair of Health Informatics}, MRI, \\ Technical University of Munich, Germany \\ \textit{GLAM - Group on Language Audio, \& Music, Imperial College London}, UK \\
\textit{https://orcid.org/0000-0002-6478-8699} 
}
}  

\maketitle

\begin{abstract}
Speech Emotion Recognition (SER) needs high computational resources to overcome the challenge of substantial annotator disagreement. Today, SER is shifting towards dimensional annotations of arousal, dominance, and valence (A/D/V). 
Instance-level measures as the L2 distance prove unsuitable for evaluating A/D/V accuracy due to non-converging consensus of annotator opinions. Concordance Correlation Coefficient (CCC) arose as an alternative measure where A/D/V predictions are evaluated to match a whole dataset’s CCC rather than L2 distances of individual audios. Recent studies have shown that Wav2Vec2 / WavLM architectures achieve today’s highest CCC. The Wav2Vec2 / WavLM family has a high computational footprint, but training small models using human annotations achieves very low CCC.
In this paper, we define a large Transformer State-of-the-Art (SotA) A/D/V model as teacher/annotator to train 5 small student models: 4 MobileNets and our proposed Wav2Small, using teacher’s A/D/V outputs instead of human annotations. The proposed teacher also sets a new SotA for valence on the MSP Podcast dataset with a CCC of 0.676. As students, we chose MobileNetV4 / MobileNetV3 , as MobileNet has been designed for fast execution times. We also propose Wav2Small -- an architecture designed for minimal parameters and RAM consumption. Wav2Small with a size of only 120\, KB when quantised for the ONNX runtime is a potential solution for A/D/V on hardware with low resources, as it has only 72\,K parameters
vs 3.12\,M parameters for MobileNet-V4-Small.
\end{abstract}


\section{INTRODUCTION}

\label{sec:intro}
\begin{table}[t]
\caption{\label{tab:macs}A/D/V model resources on 32\,bit non quantised ONNX at Intel(R) Xeon(R) Gold 6226R CPU @ 2.90GHz. All ONNX use 16kHz raw audio as input, and include all parameters, preprocessing, LogMel Spectrogram extraction.}
\vspace{.2cm}
\begin{tabular}{
  S[table-format=2.0]
  *{3}{
   S[table-format=1.0]
   S[table-format=1.0]
   S[table-format=1.0]
  }
}
\toprule
& & \multicolumn{3}{c}{5s audio input } \\
\cmidrule(lr){3-5}
& {\splitcell{Parameters (M)}} & {\splitcell{Peak \\ RSS \\ (MB)}} & {\splitcell{MAC \\ (G)}} & {\splitcell{Time CPU  \\ (ms)}} \\
\midrule
{teacher} & 483.9 & 1929 & {149} & 1089 \\ 
{WavLM \cite{goncalves24o}} & 318.6 & 1284 & {95}  & 688 \\  
{Dawn \cite{wagner23d}} & 165.3 & 697 & {55} & 372 \\  
{MobileNetV4-L} & 31.87 & 257 & {2.2} & 24 \\     
{MobileNetV4-M} & 10,38 & 94 & {1.7} &  13 \\ 
{MobileNetV3-S} & 3.14 & 22 & {0.4} & 11 \\ 
{MobileNetV4-S} & 3,12 & 36 & {0.4} & 5 \\ 
{Wav2Small} &  0,072 & 9 & {0.4} & 9 \\  
\bottomrule
\end{tabular}
\end{table}
A need for SER on low resource hardware drives us to investigate small architectures for A/D/V.
SotA A/D/V is dominated by Wav2Vec2 \cite{wagner23d} or WavLM \cite{goncalves24o}, which consist of a VGG7\footnote{VGG7 denotes 7 convolution layers, each followed by norm and activation.} feature extractor and 12 transformers layers.
VGG has a low RAM footprint from absence of skip connections, which would otherwise require storing earlier feature maps in memory \cite{vgg, carion20end}. However, the VGG7 of Wav2Vec2 \cite{wagner23d, hsu2021robust} has the same execution time as 6 transformer layers, and is prohibitively large for our aim.
In this paper, we propose an architecture for the prediction of A/D/V from speech that explores non contiguous memory reshaping to convert time frames (tokens) into the channel dimension, allowing the classifier to interpret them as `neighbourhood attention'.
For comparison, we also train $3$ MobileNetV4\footnote{Initialized with Top-1 pretrained models from https://huggingface.co/timm/ \\mobilenetv4\_conv\_aa\_large.e230\_r448\_in12k\_ft\_in1k \\mobilenetv4\_hybrid\_medium.ix\_e550\_r384\_in1k \\mobilenetv4\_conv\_small.e1200\_r224\_in1k.} \cite{qin24} (-Large, -Medium, -Small), a MobileNetV3-Small \cite{schmid23} pretrained on AudioSet \cite{gemmeke20audioset}, as well as our proposed architecture named Wav2Small.

We train the four MobileNets and Wav2Small using a custom distillation scheme defined in Sec.~\ref{sec:distill}.


\subsection{Wav2Small} \label{sec:wav2small}
The paradigm of a VGG feature extractor followed by transformer layers has shown great performance for Speech Tasks \cite{hsu2021robust}.
MobileNets \cite{howard19mobilenetv3, schmid23cpjku} do not investigate re-purposing of tokens/time-frames as convolution channels.
Transformer pretrained architectures using time-domain input are superior to LogMel (Spectrogram) input for SER, yet LogMel is still beneficial input for small architectures \cite{bagus22}.
We propose Wav2Small: a VGG7 of 13 channels followed by vectorisation of tokens into the convolution-channels dimension (Reshape operation in Listing~\ref{code}). The reshaping of tokens, mels, as channels, aids the A/D/V head (self.adv(x) in Listing~\ref{code}). Wav2Small provides 250 tokens / 1s, whereas MobileNet-V4-S provides only 16 tokens / 1s. Wav2Small can also serve as an inexpensive substitute of the feature extractor of Wav2Vec2 / WavLM / HuBERT \cite{yuan21}.

\begin{listing}[t!]
\begin{footnotesize}
\begin{Verbatim}[commandchars=\\\{\}]
\PYG{k+kn}{from} \PYG{n+nn}{torch} \PYG{k+kn}{import} \PYG{n}{nn}
\PYG{k+kn}{from} \PYG{n+nn}{torchlibrosa} \PYG{k+kn}{import}
  \PYG{n}{Spectrogram}\PYG{p}{,} \PYG{n}{LogmelFilterBank}

\PYG{k}{def} \PYG{n+nf}{conv}\PYG{p}{(}\PYG{o+ow}{in}\PYG{o}{=}\PYG{l+m+mi}{13}\PYG{p}{):}
  \PYG{k}{return} \PYG{n}{nn}\PYG{o}{.}\PYG{n}{Sequential}\PYG{p}{(}\PYG{n}{nn}\PYG{o}{.}\PYG{n}{Conv2d}\PYG{p}{(}\PYG{o+ow}{in}\PYG{p}{,}\PYG{l+m+mi}{13}\PYG{p}{,}\PYG{n}{k}\PYG{o}{=}\PYG{l+m+mi}{3}\PYG{p}{,}\PYG{n}{pad}\PYG{o}{=}\PYG{l+m+mi}{1}\PYG{p}{,}
      \PYG{n}{bias}\PYG{o}{=}\PYG{k+kc}{False}\PYG{p}{),} \PYG{n}{nn}\PYG{o}{.}\PYG{n}{BatchNorm2d}\PYG{p}{(}\PYG{l+m+mi}{13}\PYG{p}{),} \PYG{n}{nn}\PYG{o}{.}\PYG{n}{ReLU}\PYG{p}{())}

\PYG{k}{class} \PYG{n+nc}{Wav2Small}\PYG{p}{():}
  \PYG{k}{def} \PYG{n+nf+fm}{\PYGZus{}\PYGZus{}init\PYGZus{}\PYGZus{}}\PYG{p}{():}
    \PYG{n+nb+bp}{self}\PYG{o}{.}\PYG{n}{vgg7} \PYG{o}{=} \PYG{n}{nn}\PYG{o}{.}\PYG{n}{Sequential}\PYG{p}{(}
          \PYG{n}{Spectrogram}\PYG{p}{(}\PYG{n}{fft}\PYG{o}{=}\PYG{l+m+mi}{64}\PYG{p}{,} \PYG{n}{hop}\PYG{o}{=}\PYG{l+m+mi}{32}\PYG{p}{),}
          \PYG{n}{LogmelFilterBank}\PYG{p}{(}\PYG{n}{fft}\PYG{o}{=}\PYG{l+m+mi}{64}\PYG{p}{,} \PYG{n}{mels}\PYG{o}{=}\PYG{l+m+mi}{26}\PYG{p}{),}
          \PYG{n}{conv}\PYG{p}{(}\PYG{o+ow}{in}\PYG{o}{=}\PYG{l+m+mi}{1}\PYG{p}{),}
          \PYG{n}{conv}\PYG{p}{(),}
          \PYG{n}{conv}\PYG{p}{(),}
          \PYG{n}{nn}\PYG{o}{.}\PYG{n}{MaxPool2d}\PYG{p}{(}\PYG{l+m+mi}{3}\PYG{p}{,} \PYG{n}{stride}\PYG{o}{=}\PYG{l+m+mi}{2}\PYG{p}{,} \PYG{n}{pad}\PYG{o}{=}\PYG{l+m+mi}{1}\PYG{p}{),}
          \PYG{n}{conv}\PYG{p}{(),}
          \PYG{n}{conv}\PYG{p}{(),}
          \PYG{n}{conv}\PYG{p}{(),}
          \PYG{n}{conv}\PYG{p}{(),}
          \PYG{n}{nn}\PYG{o}{.}\PYG{n}{Conv2d}\PYG{p}{(}\PYG{l+m+mi}{13}\PYG{p}{,} \PYG{l+m+mi}{13}\PYG{p}{,} \PYG{n}{k}\PYG{o}{=}\PYG{l+m+mi}{1}\PYG{p}{))}
    \PYG{n+nb+bp}{self}\PYG{o}{.}\PYG{n}{lin} \PYG{o}{=} \PYG{n}{nn}\PYG{o}{.}\PYG{n}{Linear}\PYG{p}{(}\PYG{l+m+mi}{169}\PYG{p}{,} \PYG{l+m+mi}{169}\PYG{p}{)}
    \PYG{n+nb+bp}{self}\PYG{o}{.}\PYG{n}{sof} \PYG{o}{=} \PYG{n}{nn}\PYG{o}{.}\PYG{n}{Linear}\PYG{p}{(}\PYG{l+m+mi}{169}\PYG{p}{,} \PYG{l+m+mi}{169}\PYG{p}{)}
    \PYG{n+nb+bp}{self}\PYG{o}{.}\PYG{n}{adv} \PYG{o}{=} \PYG{n}{nn}\PYG{o}{.}\PYG{n}{Linear}\PYG{p}{(}\PYG{l+m+mi}{169}\PYG{p}{,} \PYG{l+m+mi}{3}\PYG{p}{)}

  \PYG{k}{def} \PYG{n+nf}{forward}\PYG{p}{(}\PYG{n}{x}\PYG{p}{):}
\PYG{+w}{      }\PYG{l+s+sd}{\PYGZsq{}\PYGZsq{}\PYGZsq{}x: (batch, time\PYGZus{}samples)\PYGZsq{}\PYGZsq{}\PYGZsq{}}
      \PYG{n}{x} \PYG{o}{\PYGZhy{}=} \PYG{n}{x}\PYG{o}{.}\PYG{n}{mean}\PYG{p}{(}\PYG{l+m+mi}{1}\PYG{p}{)}
      \PYG{n}{variance} \PYG{o}{=} \PYG{p}{(}\PYG{n}{x} \PYG{o}{*} \PYG{n}{x}\PYG{p}{)}\PYG{o}{.}\PYG{n}{mean}\PYG{p}{(}\PYG{l+m+mi}{1}\PYG{p}{)} \PYG{o}{+} \PYG{l+m+mf}{1e\PYGZhy{}7}
      \PYG{n}{x} \PYG{o}{/=} \PYG{n}{variance}\PYG{o}{.}\PYG{n}{sqrt}\PYG{p}{()}
      \PYG{n}{x} \PYG{o}{=} \PYG{n+nb+bp}{self}\PYG{o}{.}\PYG{n}{vgg7}\PYG{p}{(}\PYG{n}{x}\PYG{p}{)}
      \PYG{n}{batch}\PYG{p}{,} \PYG{n}{channel}\PYG{p}{,} \PYG{n}{token}\PYG{p}{,} \PYG{n}{mel} \PYG{o}{=} \PYG{n}{x}\PYG{o}{.}\PYG{n}{shape}
      \PYG{c+c1}{\PYGZsh{} channel is non contiguous to mel thus}
      \PYG{c+c1}{\PYGZsh{} reshapes tokens together as attention.}
      \PYG{n}{x} \PYG{o}{=} \PYG{n}{x}\PYG{o}{.}\PYG{n}{reshape}\PYG{p}{(}\PYG{n}{batch}\PYG{p}{,} \PYG{n}{token}\PYG{p}{,} \PYG{n}{channel} \PYG{o}{*} \PYG{n}{mel}\PYG{p}{)}
      \PYG{n}{x} \PYG{o}{=} \PYG{n+nb+bp}{self}\PYG{o}{.}\PYG{n}{sof}\PYG{p}{(}\PYG{n}{x}\PYG{p}{)}\PYG{o}{.}\PYG{n}{softmax}\PYG{p}{(}\PYG{l+m+mi}{1}\PYG{p}{)} \PYG{o}{*} \PYG{n+nb+bp}{self}\PYG{o}{.}\PYG{n}{lin}\PYG{p}{(}\PYG{n}{x}\PYG{p}{)}
      \PYG{n}{x} \PYG{o}{=} \PYG{n}{x}\PYG{o}{.}\PYG{n}{sum}\PYG{p}{(}\PYG{l+m+mi}{1}\PYG{p}{)} \PYG{c+c1}{\PYGZsh{} learnable pooling}
      \PYG{k}{return} \PYG{n+nb+bp}{self}\PYG{o}{.}\PYG{n}{adv}\PYG{p}{(}\PYG{n}{x}\PYG{p}{)} \PYG{c+c1}{\PYGZsh{} arousal, domin., val.}
\end{Verbatim}
\end{footnotesize}
\vspace{.4cm}
\caption{Wav2Small on PyTorch: x.reshape vectorises neighbour tokens as channel dimension before learnable pooling.\label{code}}
\end{listing} 

\section{A / D / V Distillation} \label{sec:distill}
Here, we are inspired by two publicly available SotA A/D/V models: The 12x transformer Layer Wav2Vec2 \cite{wagner23d} also referred to as Dawn and the 12x transformer layer WavLM that won the 2024 SER A/D/V Challenge \cite{goncalves24o, odyssey24} on MSP Podcast dataset \cite{lotfian2019msppodcast}.
Here, we use those 2 models to define a teacher/annotator for A/D/V.

\subsection{Teacher}
Running the WavLM and the Dawn baselines on the same audio and averaging the output A/D/V, creates a teacher model that outperforms both WavLM and Dawn.
To the best of our knowledge, this teacher defines a new SotA on MSP Podcast for valence with a CCC of 0.676, as shown in Fig.~\ref{fig:bars}. We open source the weights of the teacher\footnote{https://huggingface.co/dkounadis/wav2small}.

\subsection{Data / Augmentation}
Distillation has been applied for categorical emotions cross-lingual transfer learning \cite{ahyeon23}, as well as for A/D/V for large Wav2Vec2 / Hubert students \cite{srinivasan21}. Distillation is especially useful for A/D/V as it enables re-labeling an audio whose emotion changed inadvertently by augmentation \cite{triantafyllopoulos2022probing}.
In this paper, we opt for distillation on MSP Podcast audio and unlabelled / non-speech data.
We apply always-on mixup \cite{zhang18m}; this way, clean MSP Podcast audio is never shown
to teacher / student.
This enables long training runs without overfitting \cite{schrufer24}.
Extensive work on the most useful labels for SER \cite{chou24m} has revealed that soft labels, produced by a teacher, result in better performing students than using human annotations\cite{beyer22}; hence, for distillation-training, we use only the teacher output A/D/V as ground truth, discarding the train-split annotations.

The distillation of a student runs for 50\,M steps ($\sim$ 17 days)
with a batch size\,=\,16 (= dimension of CCC loss function),
SGD with fixed learning rate = $5\text{E}-5$, constant schedule, weight decay\,=\,0 and momentum\,=\,0,
on an Nvidia RTX A4000 GPU, using PyTorch v2.4.0.
The training audio is generated by mixing (on the fly for each batch) $0.64 \times \text{audio}_1 + 0.41 \times \text{audio}_2$, where audio$_1$ is a random 0.4\,s up to 12\,s excerpt of an audio track from one of 12 action movies
(in-house dataset containing full movies with music and effects)
and audio$_2$ is drawn from one of three buckets:
AudioSet \cite{gemmeke20audioset} sound samples,
or MSP Podcast v1.7 train speech,
or ambient urban/rural/transportation environmental background sound.
The training audio is passed through cyclic rotation (np.roll), drop of random frequencies and re-normalisation of the lost energy to the remaining non-zero frequencies (metallic speech effect) and, with probability\,$<$\,0.04, also shout augmentation \cite{tuomo08a}.
This training audio is given to the teacher and student (both 16\,kHz) and it is allowed
to overflow the [$-$1, 1] limits due to shout.


\begin{figure*}[t!]
\includegraphics[width=\linewidth]{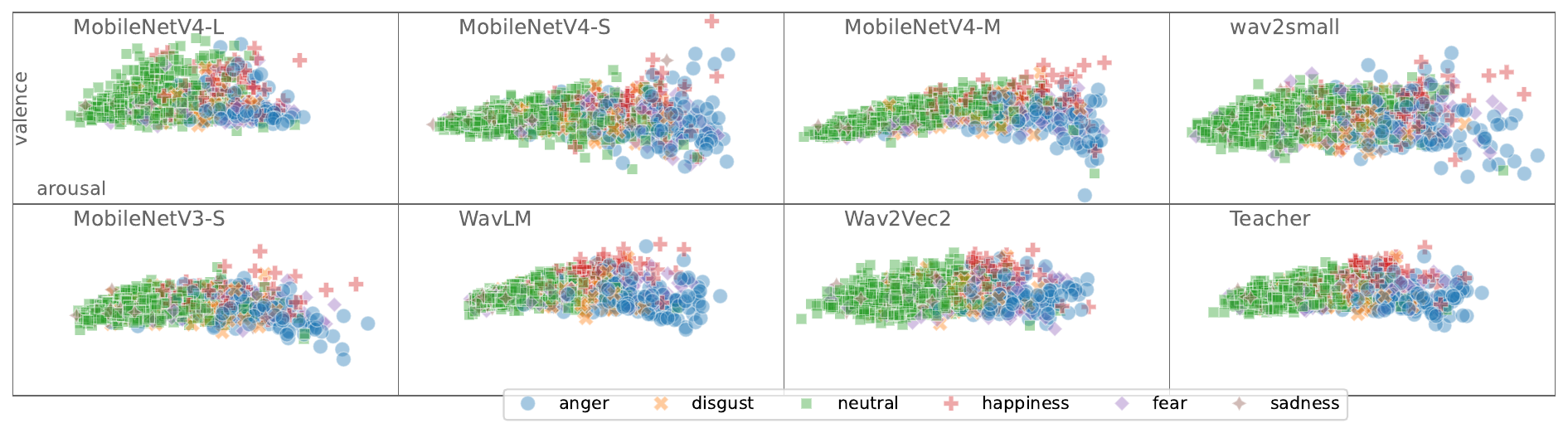}
\caption{Arousal-Valence predictions for the Crema-d \cite{cao2014crema} test set, colouring according to Crema-d ground truth categories.\label{fig:tsne}}
\end{figure*}

\begin{figure}[t!]
\vspace{-2.0cm}
\hspace{-1.0cm}
\includegraphics[scale=1.04]{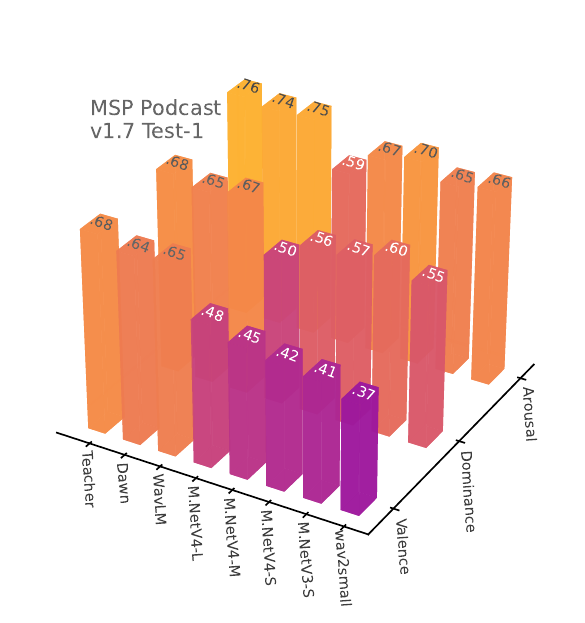}
\caption{CCC achieved by all tested models on MSP Podcast v1.7 Test1 - original human annotations.\label{fig:bars}}
\end{figure}

\subsection{Quadrant Correction Loss Function}
Annotators have higher agreement for deciding if A/D/V is positive / negative / neutral, (3D space quadrant), than deciding the exact levels of A/D/V. To reveal this information, we introduce an auxiliary loss function: If the teacher's A/D/V output falls in a different quadrant from the student's A/D/V output, we penalise the student via an extra $L1$ loss to output the A/D/V values of the teacher. This auxiliary loss vanishes if the teacher and student A/D/V quadrants agree. Then, only the CCC loss 
\cite{trigeorgis16} function remains.

\begin{figure*}[t!]
\includegraphics[width=\linewidth]{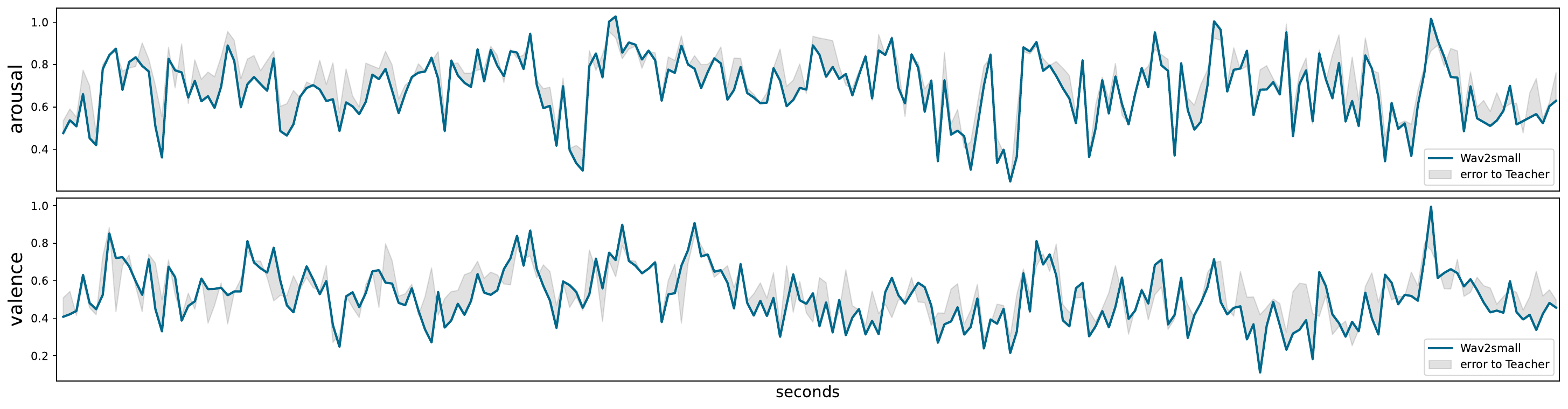}
\caption{Wav2Small discrepancy from teacher for the Japanese audio track of Harry Potter vol1. Not included for train.\label{fig:japanesse}}
\end{figure*}
\begin{figure}[t!]
\vspace{-2.4cm}
\hspace{-.84cm}
\includegraphics[scale=1.04]{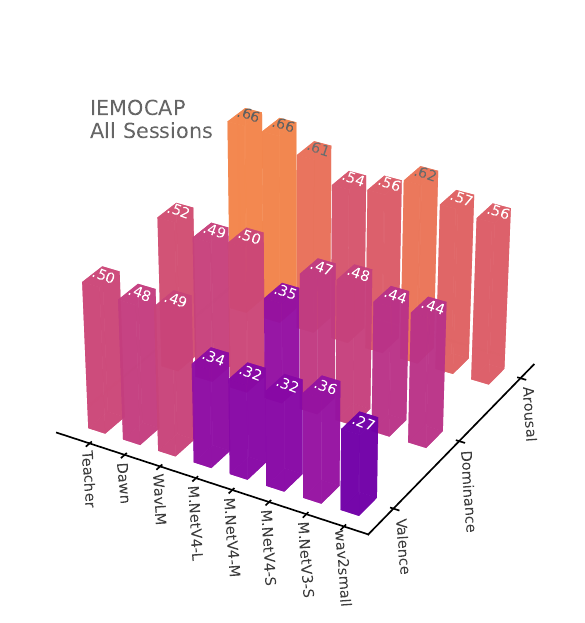}
\caption{CCC achieved by all tested models on IEMOCAP - original human annotations. Audio of all 5 Sessions has been concatenated to a single dataset. \label{fig:iemocap}}
\end{figure}
\section{Experiments} \label{sec:experiments}
Authors in \cite{wagner23d} trained the CNN14 \cite{kong17pann} of 4.9\,M parameters on MSP Podcast v1.7 human annotations. 
The CNN14 achieved valence CCC of 0.248 vs 0.64 achieved by Dawn. Now, Fig.~\ref{fig:bars} shows that MobileNetV4-S / V3-S and Wav2Small achieve valence CCC $>=$0.37, with less parameters than CNN14.

From the 5 students, MobileNets are initialised with pretrained weights and is not surprising that after distillation, they score higher than Wav2Small, for valence CCC. However, Wav2Small reaches high CCC, e.\,g., 0.66 for arousal in MSP Podcast test-1, and 0.56 for arousal in IEMOCAP \cite{busso08} used as an out of distribution dataset only for evaluation.
MobileNetV4-S attains valence CCC\,=\,0.42 and hass less execution time than MobileNetV3-S: of 5\,ms vs 11\,ms as show in Table~\ref{tab:macs}.

\subsection{Ablations}
A Frequency window of 64 FFT bands (hop\,=\,50\,\%) leads to the best CCC
for MobileNetV4-M/S, MobileNetV3-S, and Wav2Small.
However, for MobileNetV4-L we use 144 FFT bands,
because 64 FFT bands yield only 26 Mel bins, which causes a highly
asymmetric aspect ratio (if we look at LogMel as an image)
yielding empty-input in 2D convolution.
The large num FFT\,=\,144 reduces the time resolution early and therefore MobileNetV4-L
achieves only a CCC of 0.59 for arousal in Fig.~\ref{fig:bars}

All 5 students include a LogMel layer implemented in ONNX via convolutions, as ONNX is needed to run the model on low resource  hardware the Fourier/LogMel Basis is stored within ONNX as a parameter.
An extra convolution is added to MobileNet students to upscale the 1-channel output of LogMel to 3-channel RGB, needed to feed the pretrained weights.

\subsection{Vectorisation of Time as Channels}
Wav2Small opts for a shallow architecture of 13 channels producing a large amount of tokens 250/1s at 16\,KHz, at the same MACs of MobileNetV4-S.  Vectorisation of neighbouring tokens enlarges the channel dimension for the pooling (self.sof(x) in Listing~\ref{code})), creating 169 `virtual channels' for the regression head (self.adv(x) in Listing~\ref{code}).

Vectorisation/learnable pooling helps Wav2Small achieve a high CCC for arousal of 0.66 in MSP Podcast, and 0.56 in IEMOCAP (Fig.~\ref{fig:bars}/\ref{fig:iemocap}). Valence is overall, more difficult to recognise \cite{schuller2018ser}, by all models. MobileNetV4-S achieves CCC of 0.42 and Wav2Small of 0.37. Nonetheless, Fig.~\ref{fig:japanesse} shows that Wav2Small follows the teacher arousal and valence faithfully. 

\subsection{A/D/V clusters of categorical emotions}

Fig.~\ref{fig:tsne} shows that arousal / valence defines clusters that separate categorical emotions, to some extent. 
Surprisingly, the five students output more extreme valence values (attain higher vertical dispersion) in Fig.~\ref{fig:tsne}) vs Wav2Vec2 (Dawn), WavLM and the teacher. This is the last challenge of A/D/V: The fact that A/D/V models are unwilling to predict extreme valence values. Fig.~\ref{fig:tsne} shows that only MobileNetV4-S/V4-M and Wav2Small do output negative extreme valence values, although only for positive arousal. This fact exemplifies the discrepancy of perceived and felt emotions and the phenomenon that human annotators avoid very negative annotations.

\section{Outlook}
Wav2Vec2 models with $\geq$\,12 transformer layers convey linguistic cues for solving SER \cite{ioannides2023ser}, and achieve superior valence CCC ($\geq$\,0.64 in Fig.~\ref{fig:bars}) because valence is more accessible through language than acoustics \cite{schuller2018ser}; Fig.~\ref{fig:bars} shows that MobileNetV4-S achieves valence CCC\,=\,0.42 having only 3.12\,M parameters, surpassing the valence CCC\,=\,0.416 of a Wav2Vec2 of 6 transformer Layers (87.9\,M parameters) fine-tuned on human annotations \cite{wagner23d}.


In parallel to our research, interest for architectures in the 120\,KB scale emerges in the acoustic scene classification domain \cite{koutini24data};
Neural Architecture Search has birthed promising architectures for categorical emotions \cite{sun23e, rajapakshe24j}. Audio Spectrogram Transformer (AST) models \cite{kong22, koutini22passt} \cite{feng24elastic, bae23} aim to bypass the use of a VGG7 feature extractor by feeding overlapping patches of LogMel, directly as input to transformer layers. However, AST has an inherent limitation of ad-hoc selection of appropriate patches, akin to the intricacy of image patch selection in Vision Transformer \cite{dehghani23}. All those avenues open inspiration for A/D/V.
This paper focused on audio-only emotion recognition, for MSP Podcast, being one of the few A/D/V datasets collected `in the wild' paving away of acted emotions \cite{busso08, zadeh16mosi}. Fusion of text and acoustic features 
\cite{harm24t} is a potential avenue for A/D/V, although has the risk of introducing accent unfairness from focusing at words instead of sound’s tonality. Internal benchmarks of language fairness and noise tolerance \cite{derington2023a} show that Wav2Small passes $74\%$ of all tests, whereas Teacher passes $79\%$.

Dataset Distillation \cite{gutierrez24} may create teachers that provide soft-labels and synthetic-audio; overcoming the difficulty to verify that a teacher can annotate out-of-distribution speech/audio informatively.

Observer variability for various annotation schemes \cite{wood18,morgan19} reveals higher consensus for A/D/V annotations compared to annotations of 1 out of 6 emotion categories. Thus we espy A/D/V research to overshadow emotion categories.

\section{CONCLUSION}
\label{sec:conclusion}
We proposed a 72\,K parameter architecture named Wav2Small that reaches comparable A/D/V CCC scores on both MSP Podcast
and IEMOCAP datasets, while using only 9\,MB RAM compared to 36\,MB RAM used by MobileNetV4-S, while having equal MACs. Mobilenet-V4 and MobileNet-V3 reduce the time resolution to 16 : 960-dimensional tokens / 1\,s audio, whereas Wav2Small provides 250: 169-dimensional tokens / 1\,s audio, (before learnable pooling). 
Hence Wav2Small is a potential replacement of the expensive input audio extractor of transformer architectures, such as
Wav2Vec2 and WavLM. We also proposed a teacher model for distillation that obtains a new SotA on MSP Podcast achieving
a CCC of 0.676 for valence.

\bibliographystyle{IEEEbib}
\bibliography{strings,refs}

\newpage
\section*{Answers To Reviewers}

We thank the reviewers for the very detailed questions!

\section*{Reviewer \#1}

\subsubsection{Why are high computational resources needed to overcome annotator disagreement, what performance do we get if we train the smaller models from scratch without distillation?}

Training A/D/V models on ground-truth labels is difficult because augmentation invalidates the label \cite{triantafyllopoulos2022probing}:
If we add a silence in an audio of `yeah sure` it sounds as `yeah, sure` hence the perceived emotion changes. Therefore A/D/V requires either large pretrained architectures or distillation (enables augmentation of audio / relabeling by the teacher).

\subsubsection{`..VGG has a low RAM footprint from absence of skip connections..' Why do skip connections use a lot of RAM? Also, 36 MB of RAM used by MobileNetV4-S seems already small?}

Skip connections need RAM, for allocating intermediate feature tensors, e.g. layer-output tensor plus skip-connection tensor. VGG (with ReLU) operates in place, having only a single tensor in RAM.

We believe that 9MB RAM for Wav2Small is a useful reduction from the 36MB RAM of MobileNetV4-S.

\subsubsection{'The VGG7 of Wav2Vec2 has the same execution time as 6 transformer layers?} By execution time we mean latency, please see Fig.\;\ref{fig:streaming_w2v2}.

\subsubsection{`We propose an architecture for A/D/V that explores non contiguous memory reshaping as neighborhood attention?'}

The VGG7 output is a tensor (batch, channels, time-frames, mel-bins) is reshaped so that time-frames enter the "channel" dimension.

\subsubsection{The paper assumes much prior knowledge. The authors write, `MobileNets [9], [10] do not investigate re-purposing of tokens/time-frames as convolution channels..` Define MobileNets?}

A characteristic of the MobileNet family of models is reducing the input image's resolution. Wav2Small avoids down-sampling, thus has a higher resolution of time-frames than MobileNetV4-S before pooling. We discovered that Wav2Small achieves higher CCC for A/D/V for short audios $\leq1s$.

\begin{figure}[t!]
\vspace{-.4cm}
\includegraphics[width=.94\linewidth]{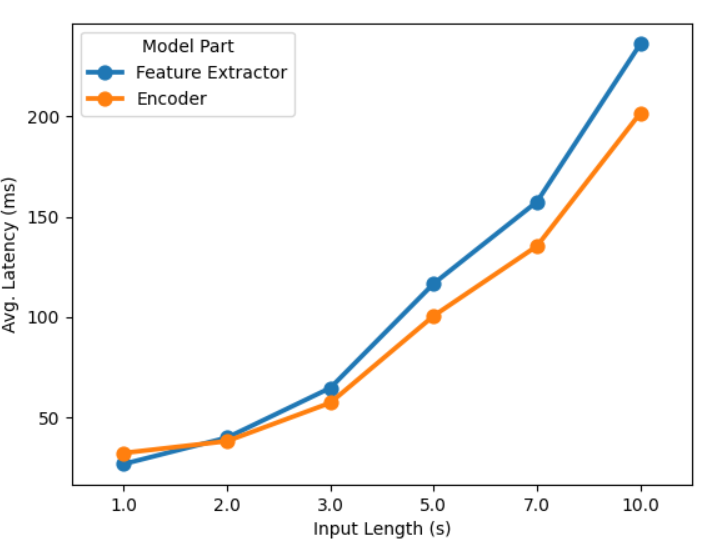}
\caption{Execution time of the 2 parts of Wav2Vec2 \cite{wagner23d, hsu2021robust}: The Feature Extractor (VGG7) and The Transformer-Encoder (prunned to 6 Transformer layers). Surprisingly, the VGG7 is slower than 6 Transformer layers for audio input $\ge 2s$. \label{fig:streaming_w2v2}}
\end{figure}

A future purpose of Wav2Small is to be used as an inexpensive feature extractor for Wav2Vec2, hence preserving high resolution = large number of time-frames, needed to feed the transformer layers. Our motivation drives from Fig.~\ref{fig:streaming_w2v2}.

\subsubsection{Typos} `For fairness` changed to 'for comparisson'. 
`Bayesian pooling' changed to 'learnable pooling'
`We propose an architecture for A/D/V.' has been changed to `We propose an architecture for the prediction of A/D/V from speech'.

\section*{Reviewer \#2}

\subsubsection{Which version of MSP PodCast is used in this work?}

Throughout this paper we used MSP-Podcast v1.7. This choice was for comparison with the experiments of [2].

\subsubsection{In Fig. 3 Was the model tested in an unseen langauge?} 
Yes. Fig.\;3. Shows that Wav2Small (72K parameters) can follow the output of its Teacher (0.4B parameters) on an unseen language.
Wav2Small has been trained as described in Section II-B.

\subsubsection{`.. MobileNetV4-S/V4-M and Wav2Small do output negative extreme valence values ..', how accurate were these predictions? Does it correlate with the overall rater consensus?}

SER datasets are labeled by approx. 10 annotators per audio and their opinions are averaged. The averaging process inevitably hides truly high/low A/D/V labels, to a more neutral value. 

\section*{Reviewer \#3}

\subsubsection{Mix-up augmentation is always on with a ratio of 0.64 for audio 1 and 0.41 for audio 2. I’m curious about the ratio setting for not summing to 1?}
Audio stored in a .wav file has values in range [-1,1]. We discovered that overflowing slightly the [-1,1] range, is a form of augmentation that improves CCC, especially for inputs of short duration $\le 2$s.

\subsubsection{The authors introduced the quadrant correction loss function. How many epochs are applied with this auxiliary loss. How do we make sure that quadrants agree in this epoch and they will keep agreeing in the following epochs?}

We always apply the quadrant correction loss. The total loss for training $=100 \times \text{CCC\_loss}() + 4 \times \text{quadrant\_correction\_loss}()$.

\subsubsection{The robustness of the proposed model might be worthy of discussion as well, e.g., noise robustness ?}

Benchmarks of language fairness and noise tolerance: https://audeering.github.io/ser-tests/ show that Wav2Small passes $74\%$ of all tests, whereas Teacher passes $79\%$.

\subsubsection{Is there any explanation why extreme predictions only appear in arousal instead of valence, Fig.\;\ref{fig:tsne}?}

A/D/V models trained with CCC loss on labels in [0,1] range, tend to output $0.5$, when it is indecisive.

\newpage

\let\thefootnote\relax\footnotetext{© 2024 IEEE. Personal use of this material is permitted. Permission from IEEE must be obtained for all other uses, in any current or future media, including reprinting/republishing this material for advertising or promotional purposes, creating new collective works, for resale or redistribution to servers or lists, or reuse of any copyrighted component of this work in other works.}

\end{document}